\documentclass{aa}  
\usepackage{graphicx}       
\usepackage{natbib}
\usepackage{longtable}
\usepackage{supertabular}
\usepackage{rotating}
\usepackage{lscape}
\usepackage{pdflscape}

\newcommand{\BCG}{{BCG}}
\newcommand{\BCDG}{{BCDG}}

\newcommand{\Ha} {H$\alpha$}

\newcommand{\HII}{\textsc{H\,ii\ }}

\newcommand{\IRAS}{{IRAS}}

\newcommand{\Mo} {$M_{\odot}$}
\newcommand{\Myr}{Myr}          
\newcommand{\NED} {{NED}}
\newcommand{\NIR} {\emph{NIR}}

\newcommand{\SFR} {{\sc Sfr}}

\newcommand{\VLT}{{VLT}}
\newcommand{\WRBUMP} {WR bump}
\newcommand{\Zo} {Z$_{\odot}$}

\newcommand{\WHa}{$W($H$\alpha)$}

\newcommand{\tableline}{\hline}

\newcommand{\nodata}{...}
\begin{document}

   \title{Massive star formation in Wolf-Rayet galaxies\thanks{Based on observations made with NOT (Nordic Optical Telescope) and INT (Isaac Newton 
Telescope) operated on the island of La Palma jointly by Denmark, Finland, Iceland, Norway and Sweden (NOT) or the Isaac Newton Group (INT) in the 
Spanish Observatorio del Roque de Los Muchachos of the Instituto de Astrof\'\i sica de Canarias. Based on observations made at the Centro 
Astron\'omico Hispano Alem\'an (CAHA) at Calar Alto, operated by the Max-Planck Institut f\"ur Astronomie and the Instituto de Astrof\'{\i}sica de 
Andaluc\'{\i}a (CSIC).}}

   \subtitle{I. Optical and \NIR\ photometric results}

   \author{\'Angel R. L\'opez-S\'anchez
          \inst{1,2}
		  \and
		  C\'esar Esteban\inst{1}
          }

   \offprints{\'Angel R. L\'opez-S\'anchez, \email{Angel.Lopez-Sanchez@csiro.au}}

\institute{Instituto de Astrof{\'\i}sica de Canarias, C/ V\'{\i}a L\'actea S/N, E-38200, La Laguna, Tenerife, Spain \and CSIRO / Australia Telescope 
National Facility, PO-BOX 76, Epping, NSW 1710, Australia}


   \date{Received: January 17, 2008; Accepted: August 3, 2008. {\bf REF: L\'opez-S\'anchez \& Esteban 2008, A\&A, 491, 131.}}

 
  \abstract
   {}
	{We have performed a comprehensive multiwavelength analysis of a sample of 20 starburst galaxies that show the presence of a substantial 
population of massive stars. The main aims are the study of the massive star formation and stellar populations in these galaxies, and the role that 
interactions with or between dwarf galaxies and/or low surface companion objects have in triggering the bursts. In this series of papers, we present 
our new optical and near-infrared photometric and spectroscopic observations, and complete with data at other wavelengths (X-ray, far-infrared, and 
radio) available in the literature. In this paper, the first in the series, we analyze the morphology, stellar population age, and star-formation 
rate of each system.}
  {We completed new deep optical and \NIR\ broad-band images, as well as the new continuum-subtracted H$\alpha$ maps, of our sample of Wolf-Rayet 
galaxies. We analyze the morphology of each system and its surroundings and quantify the photometric properties of all important objects. All data 
were corrected for both extinction and nebular emission using our spectroscopic data. The age of the most recent star-formation burst is estimated 
and compared with the age of the underlying older low-luminosity population. The \Ha-based star-formation rate, number of O7V equivalent stars, mass 
of ionized gas, and mass of the ionizing star cluster are also derived.}
     {We found interaction features in many (15 up to 20) of the analyzed objects, which were extremely evident in the majority. We checked that the 
correction for nebular emission to the broad-band filter fluxes is important in compact objects and/or with intense nebular emission to obtain 
realistic colors and compare with the predictions of evolutionary synthesis models. The estimate of the age of the most recent star-formation burst 
is derived consistently. In general, the \Ha-based star formation rate agrees with the estimates given by independent multiwavelength methods. With 
respect to the results found in individual objects, we remark the strong \Ha\ emission found in IRAS 08208+2816, UM 420, and SBS 0948+532, the 
detection of a double-nucleus in SBS 0926+606A, a possible galactic wind in Tol 9, and one (two?) nearby dwarf star-forming galaxies surrounding Tol 
1457-437.} 
   {}

\titlerunning{Massive star formation in Wolf-Rayet galaxies I: Photometric results}

\authorrunning{L\'opez-S\'anchez \& Esteban}

   \keywords{galaxies: starburst --- galaxies: interactions --- galaxies: stellar populations --- galaxies: optical/\NIR\ \& \Ha\ photometry --- 
stars: Wolf-Rayet }

   \maketitle
%

\section{Introduction}

\subsection{The nature of Wolf-Rayet galaxies}

Wolf--Rayet (WR) galaxies are a subset of emission-line and \ion{H}{ii} galaxies, whose integrated spectra show broad emission features attributed to 
the presence of WR stars, indicating that a substantial population of this type of massive star exists in the ionized cluster(s) of the 
star-formation bursts. The most massive O stars ($M\geq$ \mbox{35 \Mo}\ for \Zo) become WR stars around 2 and 3 \Myr\ after their birth, spending 
only some few hundreds of thousands of years ($t_{WR}\leq$10$^6$ yr) in this phase \citep{MaederMeynet94} until they explode as Type Ib/Ic supernovae 
\citep{vH01}. The minimum stellar mass that an O star needs to reach the WR phase and its duration is dependent on metallicity. There are two 
important broad features that reveal the presence of WR stars: the so-called blue \WRBUMP\ (between 4650--4690 \AA) and the red \WRBUMP\ (basically 
formed by the \ion{C}{iv} $\lambda$5808 emission line). The broad, stellar, \ion{He}{ii} $\lambda$4686 is the main feature of the blue \WRBUMP. The 
narrow, nebular \ion{He}{ii} $\lambda$4686 is usually associated with the presence of these massive stars, although it is rarely strong and its 
origin remains controversial \citep{Garnett91,G04}.

The detection of WR features in the spectrum of a starburst galaxy constrains the parameters that characterize the star-formation burst: the initial 
mass funtion must be extended to higher masses; the WR/O ratio is relatively large and the burst must therefore be short; and the time elapsed since 
the last starburst episode occurred must be less than a few Myr. Therefore, WR galaxies offer the opportunity to study an approximately coeval sample 
of very young starbursts \citep{SV98}.
 
The blue compact dwarf galaxy He 2-10 was the first object in which WR features were detected \citep{Allen76}.  \citet{Osterbrock82} and \citet{C91} 
introduced the concept of a WR galaxy, to be \emph{a galaxy whose integrated spectrum has detectable WR broad stellar emission lines emitted by 
unresolved stellar clusters}. \citet{KJ85} performed the first systematic search for WR features in emission-line galaxies: in their sample of 45 
extragalactic \HII regions they classified 17 as WR galaxies. \citet{KS86} and \citet{Dinerstein86} reported the first detections of the red \WRBUMP. 
\citet{C91} compiled the first WR catalogue, including 37 objects. \citet{VC92} developed the first quantitative scheme to estimate WR populations in 
starbursts using new quality data. The majority of detections of WR features have however been accidental, and have occurred in studies that cover a 
wide range of topics, from the determination of the primordial He abundance \citep{Kunth83,KJ85,ITL94,Izotov97,IT99,IT98,TIL95}, the nature of 
Seyfert galaxies \citep{heckmanetal97}, and starbursts with strong galactic winds \citep{Allen76}. \citet{GIT00} analyzed a sample of 39 objects with 
heavy element abundances ranging from \Zo/50 to 2\Zo\ and obtained global results for WR galaxies. \citet{Buckalew05} compared the properties of 
young star clusters with and without WR stars. 

The most recent catalogue of WR galaxies was compiled by \citet{SCP99} and listed 139 members, although this number has since increased 
\citep{PH00,GDHL01,BO02,CTS02,PSGD02,LTD03,Tran03,FCCG04,IPG04,PKP04,Jam04,TI05}, and these galaxies have even been detected at high $z$ 
\citep{VMCGD04}. In a study of emission-line galaxies extracted from the Sloan Digital Sky Survey \citep{York00} WR features were identified in many 
star-forming galaxies \citep{Kniazev04,Zhang07}, increasing the number of known WR galaxies to more \mbox{than 300.} 

Morphologically, WR galaxies constitute an inhomogeneous class of star-forming objects. They are detected in irregular galaxies, blue compact dwarf 
galaxies (BCDGs), spiral galaxies (or, more precisely, giant \ion{H}{ii} regions in the arms of spiral disks), luminous, merging \IRAS\ galaxies, 
active galactic nuclei (AGNs), and Seyfert 2 and low-ionization nuclear emission-line regions (LINERs) galaxies. Quoting \citet{SCP99}, the minimum 
common property of all WR galaxies is \emph{ongoing or recent star formation that has produced stars sufficiently massive to evolve to the WR stage}. 

We note that the definition of WR galaxy is \emph{dependent of the quality of the spectrum, and location and size of the aperture}. The term WR 
galaxy must therefore be used with caution. The presence of WR features in the spectrum of a starburst \emph{does not} imply that WR stars are 
present at all locations, but only that a significant population of this type of massive star exits inside the galaxy. Depending on the distance of 
the object and size of the area analyzed, the region of concern may be a single extragalactic \HII region with a few WR stars in a galaxy, a massive 
star cluster or the nucleus of a powerful starburst galaxy harbouring numerous massive stars \citep{SCP99}. The precise locations of the WR stars 
usually remain unknown, apart from for the Local Group or other nearby galaxies. The width of the extraction aperture for which the spectrum is 
extracted can sometimes be too large and the weak WR features diluted by the continuun flux. Furthermore, a starburst galaxy with several 
star-forming bursts may only show WR features in one of them.
Aperture effects and the slit position can therefore play an important role in the detection of WR features 
\citep{HGJL99,LSER04a,LSER04b,Buckalew05,LSEGR06}. 

\subsection{Aims of this paper series}

In dwarf galaxies, starburst phenomena cannot be explained by the wave-density theory because of their low masses, and an alternative mechanism must 
operate. A proposed alternative mechanism for large-scale starburst formation is gas compression by shocks due to mass loss by means of galactic 
winds and the subsequent cooling of the medium \citep{Th91,Hi00}. Other authors however proposed galaxy interactions as a massive-star formation 
triggering mechanism \citep{SS88}. Interactions appear to play a fundamental role triggering starbursts, both in spiral (Koribalski 1996; Kennicutt 
1998), and dwarf and irregular galaxies. In these cases, interactions with nearby giant galaxies are unusual \citep{CA93,TT95}, but with low 
surface-brightness galaxies \citep{WLM96,N01} or H I clouds \citep{T96,Thuan99,vZee01}. Studying a sample of WR galaxies, \citet{Mendez99} performed 
an analysis of 13 objects extracted from the catalogue of \citet{C91}, finding that 7 are clearly interacting and another 4 show features of 
interactions. For example, he found a bridge between two galaxies in Zw 0855+06 \citep{M99a}, prominent tidal tails in Mkn 8 \citep{EM99}, 
star-formation activity induced by an \ion{H}{i} cloud in Mkn 1094 \citep{M99} and an intermediate-age merger in Tol 35 \citep{ME99}. For the first 
time, these facts enabled \citet{ME00} to suggest that interactions with or between dwarf objects could be the main star-formation triggering 
mechanism in dwarf galaxies. These authors also noted that the interacting and/or merging nature of WR galaxies can be detected only when both deep, 
high-resolution images and spectra are available.

\begin{table*}[t!]
\centering
  \caption{\footnotesize{Main data of the sample of 20 WR galaxies analyzed in this work. }}
  \label{galaxias}
  \tiny
  \begin{tabular}{l@{\hspace{6pt}} c@{\hspace{6pt}}c@{\hspace{4pt}}  c@{\hspace{3pt}}   c@{\hspace{6pt}}c@{\hspace{4pt}}
r@{\hspace{6pt}}l@{\hspace{6pt}} l@{\hspace{6pt}} l@{\hspace{6pt}} }
  \noalign{\smallskip}
    \hline 
		\noalign{\smallskip}
 Galaxy  &   R.A.(2000)   &    Dec.(2000)    & $E_G (B-V)^a$ &  $m_B^b$    &  $M_B^b$   &   $d^c$        &  [O/H]$^d$ & Type$^e$ & Other \\
          &  ($h\ m\ s$)   &   ($^\circ\ '\ \arcsec$) &     &    &      &   (Mpc)    &  (dex) & & names \\  
    \hline
    \noalign{\smallskip} 
NGC 1741       &  05 01 38.4 & $-$04 15 25 & 0.051 & 13.59  & $-$20.01  &  52.5  &  8.22    & pec            & HCG 31 AC, Mkn 1089, SBS 0459-043 \\
Mkn 1087       &  04 49 44.4 &   +03 20 03 & 0.063 & 13.08  & $-$22.14  & 110.6  &  8.57$^*$& S0 pec         & II Zw 23   \\
Haro 15        &  00 48 35.9 & $-$12 43 07 & 0.023 & 13.82  & $-$20.87  &  86.6  &  8.37$^*$& (R)SB0 pec?    & Mkn 960 \\
Mkn 1199       &  07 20 28.3 &   +33 32 21 & 0.054 & 12.98  & $-$20.68  &  54.0  &  8.75$^*$& Sc \HII        & SBS 0720+335\\
Mkn 5          &  06 42 15.5 &   +75 37 33 & 0.084 & 14.83  & $-$15.57  &  12.0  &  8.07    & I? \HII        & SBS 0635+756\\
IRAS 08208+2816&  08 23 55.0 &   +28 06 14 & 0.032 & 15.10  & $-$21.29  & 190.0  &  8.42$^*$& Irr            & \nodata\\
IRAS 08339+6517&  08 38 23.2 &   +65 07 15 & 0.092 & 12.94  & $-$21.57  &  78.3  &  8.45$^*$& Pec LIRG \HII  & \nodata\\
POX 4          &  11 51 11.6 & $-$20 36 02 & 0.039 & 14.56  & $-$18.79  &  45.5  &  8.03    & \HII           & IRAS 11485-2018\\
UM 420         &  02 20 54.5 &   +00 33 24 & 0.036 & 17.32  & $-$19.55  & 237.1  &  7.95    & Compact        & SBS 0218+003\\
SBS 0926+606A  &  09 30 06.5 &   +60 26 52 & 0.031 & 16.45  & $-$17.29  &  55.9  &  7.94    & \BCG, \HII     & \nodata \\
SBS 0948+532   &  09 51 32.0 &   +52 59 36 & 0.013 & 17.93  & $-$18.43  & 187.4  &  8.03    & Sy             & \nodata \\
SBS 1054+365   &  10 57 47.0 &   +36 15 26 & 0.021 & 15.46  & $-$14.06  &   8.0  &  8.00    & NE             & \nodata \\
SBS 1211+540   &  12 14 02.5 &   +53 45 18 & 0.020 & 17.32  & $-$13.27  &  13.1  &  7.65    & \BCG           & \nodata \\
SBS 1319+579   &  13 21 10.0 &   +57 39 41 & 0.014 & 15.32  & $-$18.53  &  28.8  &  8.05$^*$& \HII\          & \nodata \\
SBS 1415+437   &  14 17 01.7 &   +43 30 13 & 0.009 & 15.32  & $-$14.52  &   9.3  &  7.58    & \BCG           & \nodata \\
III Zw 107     &  23 30 09.9 &   +25 31 58 & 0.060 & 14.36  & $-$20.14  &  79.6  &  8.23    & Im         & IV Zw 153, IRAS 23276+2515 \\
Tol 9          &  10 34 38.7 & $-$28 35 00 & 0.065 & 13.92  & $-$19.26  &  43.3  &  8.58    & E4: \HII   & 
IRAS\,10323-2819,ESO\,435-42,Tol\,1032-283\\
Tol 1457-262a  &  15 00 29.0 & $-$26 26 49 & 0.158 & 14.44  & $-$19.73  &  68.1  &  8.22$^*$& \HII       & IRAS 14575-2615, ESO 513-IG11 \\
Arp 252        &  09 44 58.6 & $-$19 43 32 & 0.049 & 16.22  & $-$19.35  & 129.8  &  8.50$^*$& Gpair pec      & ESO 566-7 + ESO 566-8\\
NGC 5253       &  13 39 55.9 & $-$31 38 24 & 0.056 & 10.09  & $-$17.92  & 4.0$^f$&  8.28$^*$& Im pec \HII    & Haro 10 \\

	\noalign{\smallskip}    
  \hline
  \end{tabular}
    \begin{flushleft}
  $^a$ Value of the Galactic extinction \citep{SFD98}.\\
  $^b$ Corrected for Galactic and internal extinction.\\
  $^c$ Except for NGC 5253, the distances were estimated from our optical spectra and correcting for Galactic Standard of Rest (see Paper II).\\ 
  $^d$ Oxygen abundance, in units of 12+log(O/H), derived in this work for each galaxy (see Paper II). If several regions were analyzed in the same 
galaxy (indicated by a star), the highest oxygen abundance derived using $T_e$ is shown.\\
  $^e$ Morphological type as indicated by \NED.\\
  $^f$ Distance obtained by \citet{Karachentsev04}.\\
  \end{flushleft}
\end{table*}

Subsequent works \citep{IPV01,VM01,VM02,Tran03} also found a relation between massive star formation and the presence of interaction signatures in 
this type of starburst. However, a systematic analysis of a significant sample of starburst galaxies containing WR stars was needed to derive more 
robust statistics and definitive results. We have therefore completed a detailed morphological, photometric, and spectroscopic study of 20 objects, 
the majority being extracted from the catalogue of WR galaxies published by \citet{SCP99}. This study combines deep optical and near-infrared (\NIR) 
broad-band and H$\alpha$ ima\-ging with optical spectroscopy (long-slit and echelle) data. Additional X-ray, far-infrared, and radio data were 
compiled from the literature. We performed a comprehensive and coherent study of all galaxies using the same reduction and analysis procedures and 
the same set of equations to determine their physical and chemical properties, with the emphasis of a global analysis of the sample. The main aims 
are to study the formation of massive stars in starburst galaxies and the role that interactions with or between dwarf galaxies and/or low surface 
brightness objects have in triggering bursts. The results of this deep analysis of local starbursts would also have an important impact on our 
knowledge about the galaxy evolution: galaxy interactions between dwarf objects should be more common at high redshifts, as hierarchical formation 
models of galaxies (i.e. Kauffmann \& White 1993; Springer et al. 2005) predict.

\subsection{Structure of the study}

We analyze our sample of WR galaxies in the following way. In this paper (Paper I), we present the photometric results derived from the optical and 
near-infrared (\NIR) broad-band and \Ha\ images.  The aims of the observations in broad-band filters are the following:
 \begin{enumerate}
  \item To analyze the stellar-component morphology of each galaxy, looking for signs of interactions (e.g. arcs, plumes, bridges, and tidal tails) 
and possible low-surface brightness companion objects. The identification and the localization of these features and external objects with respect to 
the main galaxies provides clues about their evolution, allowing us to suggest how the star-formation burst was triggered.
  \item To perform aperture photometry of each galaxy and its different regions (star-forming knots and emission-free areas) to characterize the 
stellar population that dominates the bursts and the underlying low-luminosity component. The comparison of the colors with the predictions of 
population synthesis models permits us to estimate the age of the last star-formation burst.
\end{enumerate}
We completed deep observations in narrow-band \Ha\ filters to study the extension and properties of the ionized gas. The continuum-subtracted \Ha\ 
images were used to:
   \begin{enumerate}
   \item Study the distribution of the ionized gas, and check the physical association of other surrounding star-forming objects with the main 
galaxy.
   \item Estimate the \Ha\ luminosity, which indicates the total number of ionized stars in each burst and in the galaxy, as well as the ionized gas 
mass and the star formation rate (\SFR). The total mass of the ionizing cluster can also be estimated.
   \item Calculate the \Ha\ equivalent width, which is a powerful indicator of the age of the last star-formation burst.
   \end{enumerate}
In Sect.~2 we present our observations, some details of the data reduction processes, and some useful relations. A description of the galaxies, the 
deep optical maps obtained for each system, and the photometric results for all optical and \NIR\ broad-band and \Ha\ filters are presented in 
Sect.~3. 
Some results found in the photometric analysis of our galaxy sample and its summmary are discussed in Sect.~4.

\begin{table*}[t!]
\centering
  \caption{\footnotesize{Journal of observations for broad-band optical filters. All exposure times are provided in seconds. Note that in some cases 
there are several observations per filter. Dates follow the format year/month/day.}}
  \label{observaciones_optico}
  \tiny
  \begin{tabular}{l@{\hspace{7pt}} c@{\hspace{4pt}}c@{\hspace{6pt}}c@{\hspace{6pt}} 
c@{\hspace{4pt}}c@{\hspace{6pt}}c@{\hspace{6pt}} 
c@{\hspace{4pt}}c@{\hspace{6pt}}c@{\hspace{6pt}} 
c@{\hspace{4pt}}c@{\hspace{6pt}}c@{\hspace{6pt}}}
  \noalign{\smallskip}
   \hline
	\noalign{\smallskip}
 Galaxy  &   &$U$&  &  &  $B$&    & &   $V$ & &   &  $R$   \\
       & Tel. & Date & T. exp   & Tel. & Date & T. exp     & Tel. & Date & T. exp    &  Tel. & Date & T. exp  \\
 \hline 

    \noalign{\smallskip} 
HCG 31    & NOT  & 02/10/23 & 3$\times$300 & NOT  & 02/10/23 & 3$\times$300 & NOT  & 02/10/23 & 4$\times$300 & INT  & 03/09/22 & 2$\times$200\\
Mkn 1087  & NOT$^a$ &  97/02/06 & 3$\times$300          & NOT & 03/01/20 & 3$\times$300    & 2CAHA & 00/12/19 & 3$\times$1200 & NOT & 03/01/20 &  
6$\times$300  \\
Mkn 1199  &  INT& 05/11/19 & 3$\times$300&  2CAHA & 04/11/07 & 3$\times$300 &  2CAHA  & 00/12/19  & 5$\times$400 & 2CAHA & 04/11/07 & 3$\times$300     
\\
          & NOT& 06/01/07 & 2$\times$60 & INT& 05/11/19 & 3$\times$300 & NOT& 06/01/07 & 3$\times$60  \\  
Mkn 5     &  NOT & 04/01/20 & 3$\times$300    & NOT & 04/01/20 & 3$\times$300    & NOT & 04/01/20 & 3$\times$300    & NOT & 05/04/05 & 3$\times$300     
\\
Haro 15   & INT& 05/11/19 & 4$\times$300 & INT & 05/11/19 & 3$\times$300 & 2CAHA & 00/12/19 & 3$\times$1200 &  2CAHA & 04/11/06 & 3$\times$300  \\
          & NOT& 06/01/07 & 2$\times$60  & NOT& 06/01/07 & 2$\times$60 & NOT& 06/01/07 & 2$\times$60 \\
POX 4     & NOT$^a$  &  97/02/06 & 3$\times$400    & NOT$^a$  &  97/02/06 & 3$\times$300   & NOT$^a$  &  97/02/06 & 3$\times$300  & NOT &  05/04/03 &   
3$\times$300 \\
UM 420    & INT & 05/10/06 & 3$\times$300 & INT & 05/10/06 & 3$\times$300 & INT & 05/10/06 & 3$\times$300 & 2CAHA & 04/11/06 & 3$\times$300   \\
IRAS 08208+2816& NOT & 04/01/20 & 3$\times$300    &NOT & 04/01/20 & 3$\times$300    &NOT & 04/01/20 & 3$\times$300    & NOT & 05/04/05 & 3$\times$300  
\\
          &  &  &   &    &  &   &  2CAHA & 00/12/19 & 3$\times$1200 \\
IRAS 08339+6517&     NOT  & 05/04/03 &  3$\times$300 & NOT  & 05/04/03 &  3$\times$300 & NOT  & 04/03/20 &  2$\times$300 & NOT  & 04/03/20 &   
3$\times$300 \\
SBS 0926+606A  &  NOT & 04/01/20 & 3$\times$300  & NOT & 04/01/20 & 3$\times$300  & NOT & 04/01/20 & 3$\times$300  &2CAHA & 04/11/07 &  3$\times$300   
\\
SBS 0948+532  &  NOT & 05/04/05 & 3$\times$300  & NOT & 05/04/03 & 3$\times$300  & NOT & 05/04/03 & 3$\times$300 & NOT & 05/04/03 & 3$\times$300\\
SBS 1054+365  &  NOT & 04/01/20 & 3$\times$300  &  NOT & 04/01/20 & 3$\times$300  &  NOT & 04/01/20 & 3$\times$300 & \nodata & \nodata & \nodata \\
          &  &  &   &    &  &   &  2CAHA & 00/12/19 & 3$\times$1200 \\
SBS 1211+540  &   NOT & 05/04/04 & 3$\times$300  & NOT & 05/04/04 & 3$\times$300  & NOT & 05/04/04 & 3$\times$300  & NOT & 05/04/04 & 3$\times$300   
\\
SBS 1319+579  &   NOT & 04/03/20 & 3$\times$300 & NOT & 04/03/20 & 3$\times$300 & NOT & 04/03/20 & 3$\times$300 & NOT & 05/04/03 & 3$\times$300 \\
SBS 1415+437  &  NOT & 05/04/03 & 3$\times$300 &  NOT & 05/04/03 & 3$\times$300  & NOT & 05/04/03 & 3$\times$300  & NOT & 05/04/03 & 3$\times$300    
\\
III Zw 107& INT & 05/10/06 & 3$\times$300 &  2CAHA & 04/11/07 &  3$\times$300 & 2CAHA & 04/11/07 &  3$\times$300 & INT & 05/10/06 & 3$\times$300  \\
 & & & & INT & 05/10/06 & 3$\times$300 \\
Tol 9     &  NOT & 05/04/05 & 3$\times$300 &  NOT & 05/04/05 & 3$\times$300 & 2CAHA & 00/12/19 & 3$\times$1200 &  NOT & 05/04/05 & 3$\times$300\\
Tol 1457-262a  &  NOT & 05/04/03 & 3$\times$300  & NOT & 04/03/20 & 3$\times$300  & NOT & 04/03/20 & 3$\times$300 &  NOT & 05/04/03 & 3$\times$300\\
Arp 252 &  NOT & 04/03/20 & 3$\times$300 &  NOT & 04/03/20 & 3$\times$300 &  NOT & 04/03/20 & 3$\times$300 & NOT & 05/04/04 & 3$\times$300  \\
          &  &  &   &    &  &   &  2CAHA & 00/12/19 & 3$\times$1200 \\
  \hline
  \end{tabular}
  \begin{flushleft}
  $^a$ Images published by \citet{ME00}.
  \end{flushleft}
  \end{table*}

In the second paper of this series (Paper II), we will present results derived by analyzing our intermediate-resolution spectroscopy. In the final 
paper (Paper III), we will compile the properties derived using data from other wavelengths and summarize the global analysis combining all available 
multiwavelength data. It is, so far, the most complete and exhaustive data set of this kind of galaxies, involving multiwavelength results and 
analyzed following the same procedures. We will discuss the significant role that interactions with or between dwarf galaxies play in the triggering 
of massive star formation in Wolf-Rayet galaxies.

\section{Observations}

Our photometric observations are classified into three types: broad-band optical ima\-gery (standard Johnson filters in $U$, $B$, $V$, and $R$ 
bands), narrow-band \Ha\ and adjacent continuum imagery (narrow-band filters centered at the wavelength of the \Ha\ emission line at the redshift of 
the galaxy), and broad-band \NIR\ imagery (filters in $J$, $H$ and $Ks$ bands). We describe our observations, reduction, analysis procedures, and 
present the selection criteria of our sample of WR galaxies. 

\subsection{Selection criteria of the sample galaxies}

Since we are interested in the analysis of the massive star population (Wolf-Rayet stars) in starburst galaxies, we considered the most recent 
catalogue of Wolf-Rayet galaxies \citep{SCP99} as a starting point. As we remarked in the introduction, the WR galaxy catalogue contains an 
inhomogeneous group of starbursting objects. Our analysis however is mainly focused in dwarf galaxies. Therefore, we did not consider either spirals 
galaxies or giant \HII regions within them, and considered only dwarf objects, such as apparently isolated \BCDG s and dwarf irregular galaxies that 
had peculiar morphologies in previous, shallower imaging. Finally, we chose a sample of dwarf WR galaxies that could be observed from the Northern 
Hemisphere. The only exception was NGC~5253, for which deep echelle spectrophoto\-metry using 8.2m \VLT\ was obtained (see L\'opez-S\'anchez et al. 
2007). We also chose two galaxies belonging to the Schaerer et al. (1999) catalogue that were classified as \emph{suspected} WR galaxies (Mkn 1087 
and Tol 9), to confirm the presence of massive stars within them. Finally, we also included the galaxy IRAS 08339+6517 because previous 
multiwavelength results suggested that the WR stars could still be present in its youngest star-forming bursts (see L\'opez-S\'anchez et al. 2006). 

The general properties of our galaxy sample are des\-cribed in Table~\ref{galaxias}, where we provide the equatorial coordinates, Galactic 
extinction, apparent and absolute $B$-band magnitudes, distances (assu\-ming a Hubble flow with $H_0$ = 75 km s$^{-1}$ and $q_0$ = 0.5, and 
correcting for Galactic Standard of Rest using our spectroscopic data; see Paper II), oxygen abundances (derived from our spectroscopic data; see 
Paper II), morphological type (derived from \NED), and other common names for each system.

\begin{table*}[t!]
\centering
  \caption{\footnotesize{Log of our \NIR\ observations, all completed at CST. Dates follow the format year/month/day.}}
  \label{observaciones_nir}
  \tiny
  \begin{tabular}{l cc cc cc}
  \noalign{\smallskip}
    \hline
	\noalign{\smallskip}
   &   
 \multicolumn{2}{c}{$J$} &  \multicolumn{2}{c}{$H$}  & \multicolumn{2}{c}{$K_s$} \\
 \noalign{\smallskip}	 
  Galaxy       & Date  & N$\times$Exp.Time (s)       & Date  & N$\times$Exp.Time (s)     &  Date  & N$\times$Exp.Time (s)   \\
    \tableline
    \noalign{\smallskip} 
HCG 31    & 03/02/04 & 120$\times$20 & 03/02/04 & 240$\times$10 & 03/02/04 & 360$\times$5 \\
Mkn 1087  & 02/09/24 & 120$\times$20 & 02/09/24 & 240$\times$10 & 02/09/24 & 360$\times$5 \\
Haro 15   & 02/09/24 & 120$\times$20 & 02/09/24 & 240$\times$10 & 02/09/24 & 360$\times$5  \\
Mkn 1199  & 03/02/04 & 120$\times$20 & 03/02/04 & 120$\times$10 & 03/02/04 & 360$\times$5 \\
Pox 4     & 04/02/03 & 180$\times$20 & 05/05/23 & 240$\times$10 & 05/05/23 & 240$\times$5  \\
UM 420    & 04/02/02 & 240$\times$20 & 04/02/03 & 360$\times$10 & 04/02/03 & 240$\times$5    \\

IRAS 08208+2816& 03/03/29 & 120$\times$20 & 03/03/29 & 240$\times$10 & 03/02/04 & 240$\times$5  \\
SBS 0926+606A & 03/03/26 & 180$\times$20 & 03/03/26 & 360$\times$10 & 03/03/28 & 480$\times$5  \\
SBS 1054+365  & 03/03/28 & 120$\times$20 & 03/03/28 & 360$\times$10 & 03/03/28 & 360$\times$5  \\
SBS 1319+579  & 04/02/02 & 120$\times$20 & 04/02/28 & 240$\times$10 & 04/02/28 & 360$\times$5    \\
SBS 1415+437  & 03/03/26 & 180$\times$20 & 03/03/28 & 360$\times$10 & 03/03/29 & 240$\times$5  \\
Tol 9         & 04/02/03 & 180$\times$20 & 04/02/03 & 360$\times$10 & 05/04/24 & 240$\times$5     \\
Tol 1457-262a & 04/04/18 & 120$\times$20 & 04/04/18 & 240$\times$10 & 04/04/19 & 240$\times$5    \\
Arp 252       & 04/02/01 & 180$\times$20 & 04/02/01 & 240$\times$10 & \nodata & \nodata \\
  \hline
  \end{tabular}
\end{table*}

\subsection{Optical imagery}

Images in optical wavelengths were obtained in several observing runs between the years 2000 and 2006, mainly using the 2.56m \emph{Nordical Optical 
Telescope} ({\sc Not}) located at the \emph{Roque de los Muchachos} Observatory (ORM, La Palma, Spain). However, some observations were completed at 
the 2.5m  \emph{Isaac Newton Telescope} ({\sc Int}), located at the ORM, and in the 2.2m telescope of the \emph{Centro Astron\'omico 
Hispano-Alem\'an} ({\sc Caha}) at \emph{Calar Alto} Observatory (Almer\'{\i}a, Spain). In Table~\ref{observaciones_optico}, the telescope, date, 
number of images, and exposure time for the broad-band optical observations of our galaxy sample are indicated. We observed 18 galaxies in \emph{all} 
optical broad-band filters, SBS 1054+365 was observed in all filters apart from $R$-band, and only one galaxy (NGC~5253) was not observed for which 
we adopt data from \NED.  We also used the photometric data of Mkn~1087 ($U$-band) and POX~4 ($U$, $B$ and $V$ bands) given by \citet{Mendez99}. The 
details of these observations are the following:
\begin{enumerate}
\item Observations at the 2.56m NOT. We completed three observing runs at this telescope: January-March 2004, April 2005, and April 2006. We also 
obtained data during three Spanish Service-Time nights (23 October 2002, 20 January 2003, and 7 January 2006). In all observations, the ALFOSC 
(\emph{Andalucia Faint Object Spectrograph and Camera}) instrument was used in image mode, with a CCD Loral/Lesser detector 2048 $\times$ 2048 pixel 
array, pixel size of 15 $\mu$m. The spatial resolution was 0.19$\arcsec$ pixel$^{-1}$, and the field of view was 6.3' $\times$ 6.3'.

\item Observations at the 2.2m CAHA. Two observing runs were completed at this telescope, in December 2000 and November 2004, using the CAFOS 
(\emph{Calar Alto Faint Object Spectrograph}) instrument in image mode. CAFOS was located at the Cassegrain focus of the telescope. Two different 
detectors 
were used: a CCD SITe detector with 2048 $\times$ 2048, a pixel size of 24 $\mu$m, and 0.53$\arcsec$ pixel$^{-1}$ spatial resolution during the 
observations in December 2000, and a CCD LORAL detector with 2048 $\times$ 2048, a pixel size of 15 $\mu$m, and 0.33$\arcsec$ pixel$^{-1}$ spatial
resolution for observations in November 2004. Because of the physical size of the filters, only a circular disk with a diameter of 11' is not 
vignetted by this instrument.

\item Observations at the 2.5m INT were completed on 22 September 2003 and 6 October 2005, as well as 19 November 2005 (a Spanish Service-Time night 
under non-photometric conditions). We used the \emph{Wide Field Camera} (WFC) that consists of 4 adjacent CCDs each an array of 2048 $\times$ 4096 
pixels with a pixel size of 15 $\mu$m. Located in the primary focus of the telescope, it has a spatial resolution of 0.33\arcsec\ pixel$^{-1}$, 
yielding a field of view of 11.2' $\times$ 22.4' in each chip. In our observations, only the central chip was analyzed.
\end{enumerate}
The details of the reduction process and analysis of the optical images are described in Appendix~A.

\subsection{\NIR\ imagery}

All \NIR\ observations with $J$, $H$ and $K_s$ filters were completed at the 1.5m Carlos S\'anchez Teles\-cope ({\sc Cst}), located at the 
Observatorio del Teide (Tenerife, Spain). We used the CAIN camera, which has a mosaic of 256 $\times$ 256 pixels sensitive in the 1--2.5 $\mu$m 
wavelength interval consisting of four independent chips of dimensions 128 $\times$ 128 pixels, each one controlling one quadrant of the camera. The 
physical size of each pixel is 40 $\mu$m, corresponding to  1$\arcsec$ pixel$^{-1}$ in wide field mode. The total field of view was 4'$\times$4'. 

We acquired a sequence of exposures at slightly different positions to obtain a clean sky image, following the method described in L\'opez-S\'anchez 
et al. (2004a). Table~\ref{observaciones_nir} shows the number of individual raw images obtained for each galaxy and
filter as well as the date on which they were acquired. 

We completed four observings runs at the telescope: September 2002, March 2003, February 2004, and April 2004. Additionally, we also observed on 4
February 2003 and 23 May 2005. Because of the upper limit in declination of the CST (65$^\circ$), Mkn~5, and IRAS~08339+6517 were not observed. The 
starburst galaxy NGC 5253 was not observed because it is a southern object. Three galaxies (III~Zw~107, SBS~0948+532 and SBS~1211+540) were not 
observed because of several technical problems and/or bad weather conditions. Therefore, only 14 galaxies of our sample were observed in \NIR\ using 
CST, 13 of them using all filters (Arp~252 was not observed in $K_s$). For those objects, we did not acquire new \NIR\ data, but used instead results 
given by the \emph{Two Micron All Sky Survey} (2MASS, see Cutri et al. 2000; Jarrett et al. 2000) project. The details of the reduction process and 
analysis of the NIR images are described in Appendix~B.

\begin{table*}[t!]
\centering
  \caption{\footnotesize{Log of the \Ha\ observations. Times are indicated in seconds. Dates follow the format year/month/day.}}
  \label{observaciones_ha}
  \tiny
  \begin{tabular}{l c@{\hspace{7pt}}c@{\hspace{7pt}}    c@{\hspace{7pt}}c@{\hspace{7pt}}c@{\hspace{7pt}}    
c@{\hspace{7pt}}c@{\hspace{7pt}}c@{\hspace{7pt}}c@{\hspace{7pt}}}
    \hline
	\noalign{\smallskip}
Galaxy    & Telescope &  Date  &  \Ha\ Filter &  Time  &  $K$      &    \Ha\ cont. Filter     & Time &  $K$    & seeing$^a$ ($\arcsec$)\\
    \noalign{\smallskip}
	\hline
    \noalign{\smallskip} 
HCG 31         &2.2CAHA&04/11/06& 667/8   &  4$\times$300 &  1.44 & 683/9  & 1$\times$300 & 1.37 & 1.1   \\
Mkn 1087       &2.2CAHA&04/11/06& 674/7   &  4$\times$300 &  1.21 & 727/16 & 2$\times$300 & 1.23 & 1.3   \\
Haro 15        &2.2CAHA&04/11/06& 667/8   &  3$\times$300 &  1.65 & 683/9  & 1$\times$300 & 1.73 & 1.5 \\
Mkn 1199       &2.2CAHA &00/12/20& 667/8  &  3$\times$600  & 1.03 & 613/12 & 3$\times$600 & 1.00 & 2.2  \\
Mkn 5          & NOT & 05/04/04 &  IAC-20 &  3$\times$300 &  1.54 & IAC-36 & 2$\times$300 & 1.55 & 0.8 \\
POX 4$^b$      & NOT & 97/02/04 &  IAC-24 &  3$\times$900 &  --   & NOT-21 & 3$\times$600 & --   & 1.2  \\
UM 420         &2.2CAHA &04/11/06& 696/15 &  3$\times$300 &  1.26 & 667/8  & 1$\times$300 & 1.30 & 1.0  \\
IRAS 08208+2816& NOT & 04/01/20 &  IAC-36  &  3$\times$300 &  1.02 & NOT-21 & 3$\times$300 & 1.00 & 0.6\\
IRAS 08339+6517& NOT & 04/03/20 &  IAC-19  &  3$\times$300 &  1.25 & IAC-20 & 2$\times$300 & 1.26 & 0.6 \\
SBS 0926+606A  & 2.2CAHA&04/11/07& 667/8   &  3$\times$300 &  1.12 & 683/9  & 1$\times$300 & 1.10 & 1.4 \\
SBS 0948+532   & NOT & 05/04/05 &  IAC-36  &  3$\times$300 &  1.10 & IAC-19 & 2$\times$300 & 1.11 & 1.4 \\
SBS 1054+365   & NOT & 04/01/20 &  NOT-21  &  4$\times$300 &  1.06 & --     & --           & --   & --  \\  
               & NOT & 04/03/20 &   --     &      --       &   --  & IAC-36 & 2$\times$300 & 1.04 & 0.7  \\
SBS 1211+540   & NOT & 05/04/04 &  IAC-20  &  3$\times$300 &  1.16 & IAC-36 & 2$\times$300 & 1.19 & 0.6 \\
SBS 1319+579   & NOT & 04/03/20 &  IAC-12  &  3$\times$300 &  1.18 & IAC-36 & 2$\times$300 & 1.20 & 0.7 \\
               & NOT & 05/04/03 &  IAC-12  &  4$\times$300 &  1.15 & IAC-36 & 2$\times$300 & 1.14 & 0.8 \\
SBS 1415+437   & NOT & 05/04/03 &  IAC-20  &  3$\times$300 &  1.06 & IAC-36 & 2$\times$300 & 1.08 & 0.6 \\
III Zw 107     &2.2CAHA&04/11/06& 667/8   &  3$\times$300 &   1.03 & 683/9  & 1$\times$300 & 1.05 & 1.0 \\
Tol 9          & NOT & 06/04/26 &  IAC-24  &  3$\times$900 &  1.85 & IAC-36 & 3$\times$300 & 2.00 & 0.9 \\
Tol 1457-262a  & NOT & 04/03/20 &  IAC-20 &  3$\times$300 &  1.80 & IAC-19 & 3$\times$300 & 1.85 & 1.0 \\
Arp 252      & NOT & 04/01/20 & IAC-39  &  4$\times$300 & 1.52  & IAC-36 & 3$\times$300 & 1.55 &   0.7  \\
	\noalign{\smallskip}    
  \hline
  \end{tabular}
      \begin{flushleft}
  $^a$ The worst value for the seeing is indicated.\\
  $^b$ Images from \citet{ME00}.\\
  \end{flushleft}
\end{table*}

\subsection{H$\alpha$ imagery}

\Ha\ and adjacent continuum images were obtained during the same runs used for the observation of the broad-band images, and therefore using the same
telescopes and instrumentation. We chose adequate narrow-band filters (with a FWHM of $\sim$50 \AA) to detect the redshifted \Ha\ $\lambda$6562.82
emission line taking into account the recession velocity of the object given by the NASA/IPAC Extragalactic Database (NED) and/or our optical
spectra. We obtained \Ha\ images for all galaxies in our sample apart from POX~4 and NGC 5253, for which we used the results provided by 
\citet{Mendez99} and \citet{Meurer06}, respectively. Table~\ref{observaciones_ha} compiles all the data (date, telescope, filters, exposure time, 
airmasses, and worst seeing) concerning our \Ha\ observations. The quality of these observations is remarkable: the worst seeing of the \Ha\ images 
for 9 up to 19 (47 \%) of the galaxies is lower than 1$\arcsec$. The details of the reduction process and analysis of the \Ha\ images are described 
in Appendix~C.

\subsection{Stellar populations}

For all galaxies and knots, we compared our optical/\NIR\ colors (corrected for extinction and emission of the ionized gas) with the predictions 
given by three different popu\-lation synthesis models, STARBURST99 \citep{L99}, PEGASE.2 \citep{PEGASE97}, and \citet{BC03}, to estimate the age of 
the dominant stellar population of the galaxies, the star-forming regions, and the underlying stellar component. We selected these models because 
while, the first are based on Geneva tracks, the other two use Padua isochrones \citep{padua94} in which thermally pulsing asymptotic giant branch 
(TP-AGB) phases are included. We assumed an instantaneous burst with a Salpeter IMF, a total mass of $10^6$ \Mo, and a metallicity of $Z/Z_{\odot}$ = 
0.2, 0.4 and 1 (chosen in function of the oxygen abundance of the galaxy derived from our spectroscopic data, see Paper II) for all models. Since 
these models are optimized to study the youngest stellar populations within the galaxies, ages above 500 \Myr\ cannot be measured reliably, but their 
values are useful for discriminating between young ($\leq$25 \Myr), intermediate (100--300 \Myr), and old ($>$500 \Myr) stellar populations (see 
\cite{LSEGR06} for details of the method). We used the $W$(\Ha) of all analyzed star-forming knots to estimate the age of their most recent 
starbursting episode comparing with the predictions given by the STARBURST99 \citep{L99} models (last column in Table~7), which have a far smaller 
error (between 0.1 and 0.5 \Myr) than the ages derived using broad-band colors (typically, between 2 and 5 \Myr\ for young stellar populations). In 
Paper II we will show that the ages derived from \WHa\ are in good agreement with those derived from the spectroscopic data. As we conclude in 
Sect.~4 and Paper III, a proper estimate of the stellar population age for  this type of galaxy using broad-band filters is only obtained when bursts 
and underlying components are independently considered.

\section{Results}

CHECK the full paper in the A\&A webpage:\\
{\bf http://www.aanda.org/index.php?option=article\&access=doi\&doi=10.1051/0004-6361:200809409}\\
or in the personal webpage of the PI: \\
{\bf http://www.atnf.csiro.au/people/Angel.Lopez-Sanchez/papers/9409WRG1.pdf} \\
False color pictures of the galaxies can be found in:\\
{\bf http://www.atnf.csiro.au/people/Angel.Lopez-Sanchez/pictures.html} 

\section{Summary}

CHECK the full paper.

\section{Appendices}

CHECK the full paper.

\begin{acknowledgements}

We are indebted to Ver\'onica Melo, Ismael Mart\'{\i}nez-Delgado, Mercedes L\'opez-Morales, Mar\'{\i}a Jes\'us Ar\'evalo, Alfred Rosenberg and David 
Mart\'{\i}nez-Delgado for share their observing time with us. \'A.R. L-S thanks C.E. (his PhD supervisor) for all the help and very valuable 
explanations, talks and discussions along these years. He also acknow\-ledges Jorge Garc\'{\i}a-Rojas, Sergio Sim\'on-D\'{\i}az and Jos\'e Caballero 
for their help and friendship during his PhD, extending this acknowledge to all people at Instituto de Astrof\'{\i}sica de Canarias (Spain). 
\'A.R. L-S. \emph{deeply} thanks to Universidad de La Laguna (Tenerife, Spain) for force him to translate his PhD thesis from English to Spanish; he 
had to translate it from Spanish to English to complete this publication. The authors are very grateful to A\&A language editor, Claire Halliday, for 
her kind revision of the manuscript. This work has been partially funded by the Spanish Ministerio de Ciencia y Tecnolog\'{\i}a (MCyT) under project 
AYA2004-07466. 
This research has made use of the NASA/IPAC Extragalactic Database (NED) which is operated by the Jet Propulsion Laboratory, California Institute of 
Technology, under contract with the National Aeronautics and Space Administration. This publication makes use of data products from the Two Micron 
All Sky Survey, which is a joint project of the University of Massachusetts and the Infrared Processing and Analysis Center/California Institute of 
Technology, funded by the National Aeronautics and Space Administration and the National Science Foundation.

\end{acknowledgements}


\end{document}